\documentclass[twocolumn,showpacs,preprintnumbers,amsmath,amssymb,aps,prl]{revtex4}
\usepackage{amsfonts,amssymb,amsmath}
\usepackage[dvips]{color}
\usepackage[latin1]{inputenc}
\usepackage[dvips]{graphicx}

\newcommand{\bc}{\begin{center}}
\newcommand{\ec}{\end{center}}

\newcommand{\ud}{\textrm{d}}

\begin{document}

\title{Gauge theory picture of an ordering transition in a dimer model}

\author{D.\ Charrier}
\author{F.\ Alet}
\author{P.\ Pujol}
\affiliation{Laboratoire de Physique Th\'eorique - IRSAMC, UPS and CNRS, Universit\'e de Toulouse, F-31062 Toulouse, France}

\date{\today}

\pacs{71.10.Pm, 74.20.Mn, 74.81.-g, 75.40.Mg}

\begin{abstract}
We study a phase transition in a 3D lattice gauge theory, a "coarse-grained"
version of a classical dimer model. Duality arguments
indicate that the dimer lattice theory should be dual to a XY model coupled to a
gauge field with geometric frustration. The transition between a
Coulomb phase with dipolar correlations and a long range ordered columnar
phase is understood in terms of a Higgs mechanism. Monte Carlo simulations
of the dual model indicate a continuous transition with exponents close but
apparently different from those of the 3d XY model. The continuous nature of
the transition is confirmed by a flowgram analysis.
\end{abstract}
\maketitle
Realizing that seemingly unrelated phenomena obey the same rules is
certainly one of the most fascinating aspects of physics. In the context of phase transitions between different states of
matter, this issue of universality is often addressed within
Landau-Ginzburg-Wilson (LGW) theory, where one derives an action in powers
of the order parameter describing a spontaneous symmetry breaking.
Recently, new kinds of \textit{unconventional} phase transitions which do
not easily fit in the LGW framework have been discussed in the
domain of strongly correlated systems. A first example is the
quantum phase transition arising in some frustrated antiferromagnets from a
N\'eel state with antiferromagnetic order (AF) to a valence bond solid (VBS)
state which breaks lattice symmetries~\cite{Senthil}. Another case is the classical
interacting dimer system on the cubic lattice~\cite{Alet}, which displays a
continuous transition from an ordered phase where the dimers align in columns
to a Coulomb phase with dipolar dimer correlations. In all cases, two important issues have to been considered: first, one has to make sure that the transition is indeed critical and not first-order. Then, one must seek for alternative (``non-LGW'') descriptions of the
phase transition. For instance, the possibility of a continuous AF-VBS transition has been proposed to be understood in terms of spinon deconfinement~\cite{Senthil}. However, this deconfined quantum criticality scenario is confronted with recent simulations favoring a very weak first-order driven process~\cite{Kuklovnew}.

In this paper, we focus on the effective description of the phase
transition in the classical dimer model. There, a gauge field
arises naturally in the Coulomb phase \cite{HKMS, Alet}. One generally introduces the lattice electric field:
$E_i(\textbf{r})=(-1)^{\textbf{r}}n_i(\textbf{r})$ with $\textbf{n}$  being
the dimer occupation number. To capture the hardcore nature of the dimers,
a divergence constraint is obeyed: $\mathbf{\nabla} \cdot \textbf{E} = \pm 1$. In the Coulomb
phase, the long wavelength properties of the system are described by the
coarse-grained action: $S = \frac{K}{2}\int \textbf{E}^2 \ud^3r$. 
Due to the energy form of the microscopic model, the dimer system is driven
at low temperatures to a columnar phase with broken lattice
symmetries. Using some duality transformations, we argue in this paper that
this transition can be understood in terms of a Higgs mechanism. At the
critical point, the system is described by a field theory closely linked to
the one appearing in studies of deconfined quantum criticality. The same
scenario has been recently proposed by mapping the classical
dimer model to a quantum 2d bosonic model~\cite{Powell2}. We study
via Monte Carlo (MC) simulations the transition of the effective dual model
between Coulomb and columnar phases to recover its continuous nature.

We start by considering a ``coarse-grained'' model where the electric field on
the lattice can take all integer values, which generalizes the
microscopic case. Nevertheless, we retain the divergence constraint
$\mathbf{\nabla} \cdot \textbf{E} = \pm 1$. Our starting point is the following action on the lattice:  
\begin{eqnarray*}
Z &=&\int_{-\pi }^{\pi }\left[ D\theta \right] \sum_{E_{i}(\mathbf{r)}
=-\infty }^{+\infty }e^{-S_{\mathrm{Coulomb}}-S_{\mathrm{const}}} \nonumber \\
S_{\mathrm{Coulomb}} &=&\frac{J}{2}\sum_{\mathbf{r}}E_{i}^{2}\left( \mathbf{r}\right) \nonumber \\
S_{\mathrm{cons}} &=&i\sum_{\mathbf{r}}\theta (\mathbf{r)}\left( \nabla _{i}E_{i}^{0}(%
\mathbf{r)}-\nabla _{i}E_{i}\left( \mathbf{r}\right) \right).
\end{eqnarray*}
with $E_{i}^{0}(\mathbf{r)=}\frac{(-1)^{\mathbf{r}}}{6}e_{i}$ and the
angular field $\theta$ is a Lagrange multiplier ensuring the divergence
constraint. A real-valued field is then introduced via the
Poisson formula:
\begin{eqnarray*}
\sum_{E_{i}(\mathbf{r})=-\infty}^{+\infty}e^{-S\left[ E_{i}\left( \mathbf{r
}\right) \right] } \rightarrow &&  \\
\sum_{u_{i}\left( \mathbf{r}\right) =-\infty }^{+\infty
}\int_{-\infty }^{+\infty }\left[ D\mathbf{E}\right] & \exp\left( -S\left[
E_{i}\left( \mathbf{r}\right) \right] -\sum_{\mathbf{r}}2i\pi E_i u_{i}\left( 
\mathbf{r}\right) \right). 
\end{eqnarray*}
Performing a duality transformation~\cite{Motrunich1}, we introduce the dual current $q_i=\epsilon _{ijk} \nabla_j u_k$ and the gauge field $\textbf{A}$ defined by $\mathbf{B}=\frac{1}{2\pi }\epsilon _{ijk} \nabla_{j}A_{k}$ where $B_{i}=E_{i}-E_{i}^{0}$. We also define the static vector $\textbf{X}^0$ by $E_i^0 = \frac{1}{2 \pi}\epsilon _{ijk} \nabla_{j}X_{k}^{0}$. The circulation of $\textbf{X}^0$ is equal to $\frac{\pi}{3}$ \textit{modulo}  $ 2 \pi$ on each dual plaquette. The dual theory corresponds now to a magnetic field coupled to the currents \textbf{q} with a static frustration field $\textbf{X}^0$. Adding a fugacity term for the currents: $S_f = \frac{1}{2 \lambda_m}\sum (q_i)^{2}$ , one can easily perform the integration over the \textbf{q} variables and find: 
\begin{eqnarray}
S\left[A_{i},\chi \right] &=& \frac{\mathcal{K}}{2} \sum_{\mathbf{r}^{\ast
}}\left( \nabla _{i}A_{j}-\nabla _{j}A_{i}\right) ^{2} \nonumber \\
&-&\lambda _{m} \sum_{\mathbf{r}^{\ast
}}\cos \left( \nabla _{k}\chi -A_{k}-X_{k}^{0}\right),
\label{lattice}
\end{eqnarray}
with the field $\chi$ resulting from the conservation of the dual
currents. The different steps above were already discussed for related 3D quantum spin models~\cite{Motrunich1,Motrunich2}.
We are left with a dual theory of one matter field interacting with a non
compact gauge field with geometric frustration. Note that the
non-compactness of the field originates from the absence of monomers in the
original dimer model. We can now discuss the possible phases encountered in
this dual model. For small $\lambda_m$, the gauge field is essentially free
and exhibits dipolar correlations. When $\lambda_m$ increases, the matter
field condenses and gaps the gauge field by the Higgs mechanism. However,
this condensation is constrained by the frustration vector $\textbf{X}^0$. 

To find how the matter field condenses, one standard
possibility~\cite{Motrunich1, Motrunich2, Blankschtein et al.} is to consider the soft
spin version of Eq.~\ref{lattice}. The kinetic energy turns out to have a
two-dimensional manifold of minima, giving rise to an effective
Ginzburg-Landau action function of the two complex matter fields
$\mathbf{\phi} = \left( \phi_1, \phi_2 \right)$~\cite{Motrunich1}: 
\begin{equation}
S  = \int \ud^{3}r  \left( |(\nabla _{i} - iA_{i}){\mathbf{\phi} }|^{2} +  U(\phi ) + \frac{K}{2}(\nabla \times
\mathbf{A})^2 \right),
\label{theorie3} 
\end{equation}
with $U(\phi)=r|\mathbf{\phi} |^{2}+u|\mathbf{\phi}
|^{4}+v_{8}I_{8}({\mathbf{\phi} })$ and where the exact form of the 8th order term $I_{8}$
has been derived in Ref.~\onlinecite{Motrunich1} for a quantum-mechanical 
model in $3+1$ dimensions. 

The criticality of field theories such as Eq.~\ref{theorie3} has been of great
interest in recent years. In $(2+1)$ dimensions, it is in principle related to
the AF-VBS transition at the
deconfined quantum critical point~\cite{Senthil}, up to the $I_8$ term. The
connection between the original dimer model and the effective
action Eq.~\ref{theorie3} was recently derived independently~\cite{Powell2}.
Note that the presence of the
frustration vector $\textbf{X}^0$ in Eq.~\ref{lattice} is of crucial importance for
the analysis above. The non-frustrated model, being dual to a XY theory, is known to present an inverted XY transition which therefore
lies in the 3D XY universality class~\cite{Dasgupta-Bartolomew}. Its
associated field theory corresponds to a \textit{single} matter field interacting with a gauge field. We therefore expect a 
different behavior with and without frustration. 

Theoretically, the $\epsilon$ expansion is of no help to characterize the
transition as it predicts a first order transition~\cite{Halperin-Lubensky} for both one component and two component matter
field model. Several high performance simulations~\cite{Sudbofirst,
Kuklov} on easy-plane lattice versions of two matter fields models have shown evidence for a weak first
order transition, in contrast with the continuous transition predicted by the
deconfined criticality scenario. Recent simulations on $SU(2)$ versions are
controversial, with some pointing towards a continuous
transition~\cite{Motrunichnew} while sophisticated analyses favor
a first-order process~\cite{Kuklovnew}.

There are several caveats when considering the two-matter fields model. The first problem comes from the inherent
approximations performed when taking the soft spin version of
Eq.~\ref{lattice}, and then going back to
the lattice for the numerical simulations. Secondly, simulating
Eq.~\ref{theorie3} is more complicated than Eq.~\ref{lattice} as we have
more degrees of freedom with the two-matter fields. Finally and most importantly, the connection is
lost with the original dimer model, our main motivation.
In some cases, it might be simpler to go one step back and simulate 
an intermediately-derived field theory such as Eq.~\ref{lattice}, when available. This has the
advantage of being at the same time closer to the microscopic dimer
model (and therefore making it possible to use microscopic observables such as the
columnar order parameter) as well as to a gauge-theory description
(which is necessary to invoke the Higgs mechanism). In this paper, we adopt
this strategy and present results of a MC simulation performed on
Eq.~\ref{lattice} which shows a transition between a columnar and a
dipolar phase. The transition is continuous with exponents found close but possibly
different from those of the
3d XY universality class. We used the Metropolis algorithm on cubic lattices
of size $L^3$ with periodic boundary conditions up to $L=64$, taking $\lambda_m = \beta = 1/T$.

\begin{figure}
{\vspace*{-0.3cm}}\includegraphics*[width=7cm]{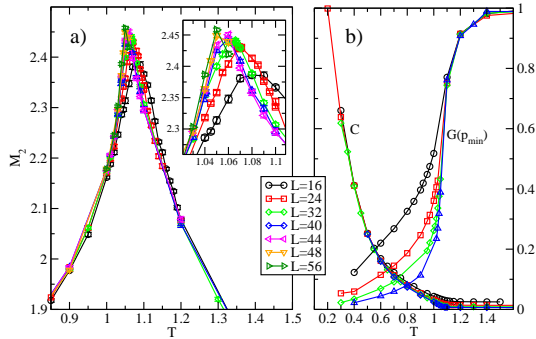}{\vspace*{-0.3cm}}
%{\vspace*{-0.3cm}}\includegraphics*[width=6cm]{Col-Mn2.eps}{\vspace*{-0.3cm}}
\caption {(color online) a) Second moment $M_2$ versus $T$ for different system
sizes. Inset: Zoom on the peak. b) Columnar order parameter $C$ and transverse magnetic field
correlator $G(\textbf{p}_\mathrm{min})$ as a function of $T$ for different
system sizes.}
\label{fig:CV}
\end{figure}

We first take $\mathcal{K} = 1$ and  present thermodynamic results. We checked that the probability
distribution of the action reveals no double peak structure, supporting the scenario of a second order transition. The second moment
of the action $M_2 = \langle\left( S-\langle S \rangle\right) ^2
\rangle/L^2$ displays a peak at the transition point (see
Fig.~\ref{fig:CV}a). The height of the peak grows slowly up to $L=32$ and
then converges within error bars for larger systems. The critical temperature $T_c^{M_2} = 1.05(1)$  is estimated 
from the position of the maximum. The fact that $M_2$ converges at the
transition indicates a critical
exponent $\alpha$ negative, as in the 3d
XY universality class. We note that $M_2$ appears to converge more rapidly
with system size than for the non-frustrated model (data not shown), similar
to the situation found in Ref.~\onlinecite{Motrunichnew}. This suggests an exponent $\alpha$ larger (in absolute
value) than for the 3d XY universality class  although it is difficult to give a
precise value.

\begin{figure}
{\vspace*{-0.3cm}}
\includegraphics*[width=7cm]{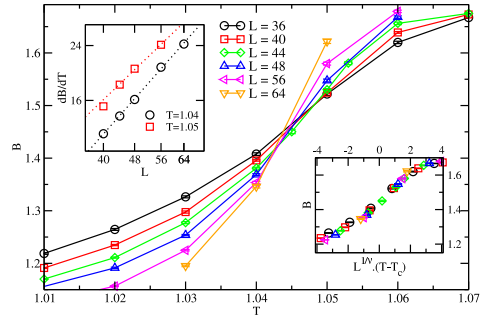}{\vspace*{-0.3cm}}
\caption{(color online) Binder cumulant of the columnar parameter versus $T$. Left
inset: Divergence of the derivative $dB/dT$ close to $T_c$ versus system size (log-log scale). Right inset: Binder cumulant scaling collapse.}
\label{fig:Binder}
\end{figure}

We now consider the low-temperature phase. In this lattice gauge theory, a gauge-invariant order
parameter related to lattice symmetry breaking can be defined. In the original 
microscopic model, dimers order on columns as temperature is
lowered. Equivalently in the coarse grained model, the magnetic field
\textbf{B} arranges in staggered flux lines in one particular
direction. We therefore define the local order parameter
$c_i(\textbf{r}) = (-1)^{\textbf{r}-r_i}B_i(\textbf{r})$ and the 
associated global parameter:
\begin{equation*}
C = \frac{1}{L^3} \Vert \textbf{C} \Vert = \frac{1}{L^3} \Vert \sum_{\textbf{r}} \textbf{c}(\textbf{r}) \Vert.
\end{equation*}
At low $T$, we expect one of the component of the vector \textbf{C} to be non-zero,
resulting in a finite expectation value. The original columnar states of the
dimer model are represented by the configurations with: $\vec{C} =
\left\lbrace \pm C,0,0 \right\rbrace, \left\lbrace 0,\pm C,0 \right\rbrace , \left\lbrace 0,0,\pm C \right\rbrace$. 
In Fig.~\ref{fig:CV}b, $C$ is observed to vanish at high T, to take a non-zero value as
$T$ decreases and to finally behave as $1/T$ at low $T$ as can be
understood from the equations of motion. To locate the critical point, we
measure the Binder cumulant of the order parameter $B = \langle C^4 \rangle
/  \langle C^2 \rangle ^2$, which admits a crossing point for different
systems sizes (see Fig.~\ref{fig:Binder}). This is characteristic of a
second order transition and leads to an estimate $T_c^{col}
=1.044(5)$. Assuming the standard scaling form $B = f (L^{1/\nu}(T-T_c))$,
the derivative $dB/dT$ should scale as $L^{1/\nu}$ at criticality. We have
measured this quantity thermodynamically, and display its scaling in the
left inset of Fig.~\ref{fig:Binder} for the temperatures $T=1.04$ and
$T=1.05$  around the estimated $T_c$. Fits to a power-law form at these two
$T$ allow to bound $0.70<\nu<0.82$. The other inset shows the best data collapse
of the curves according to the scaling form with $\nu =0.73$ and $T_c=1.044$. We find that
acceptable data collapses can also be obtained for values in the
rather broad range $\nu=0.65-0.85$.

\begin{figure}
{\vspace*{-0.3cm}}\includegraphics[width=7cm]{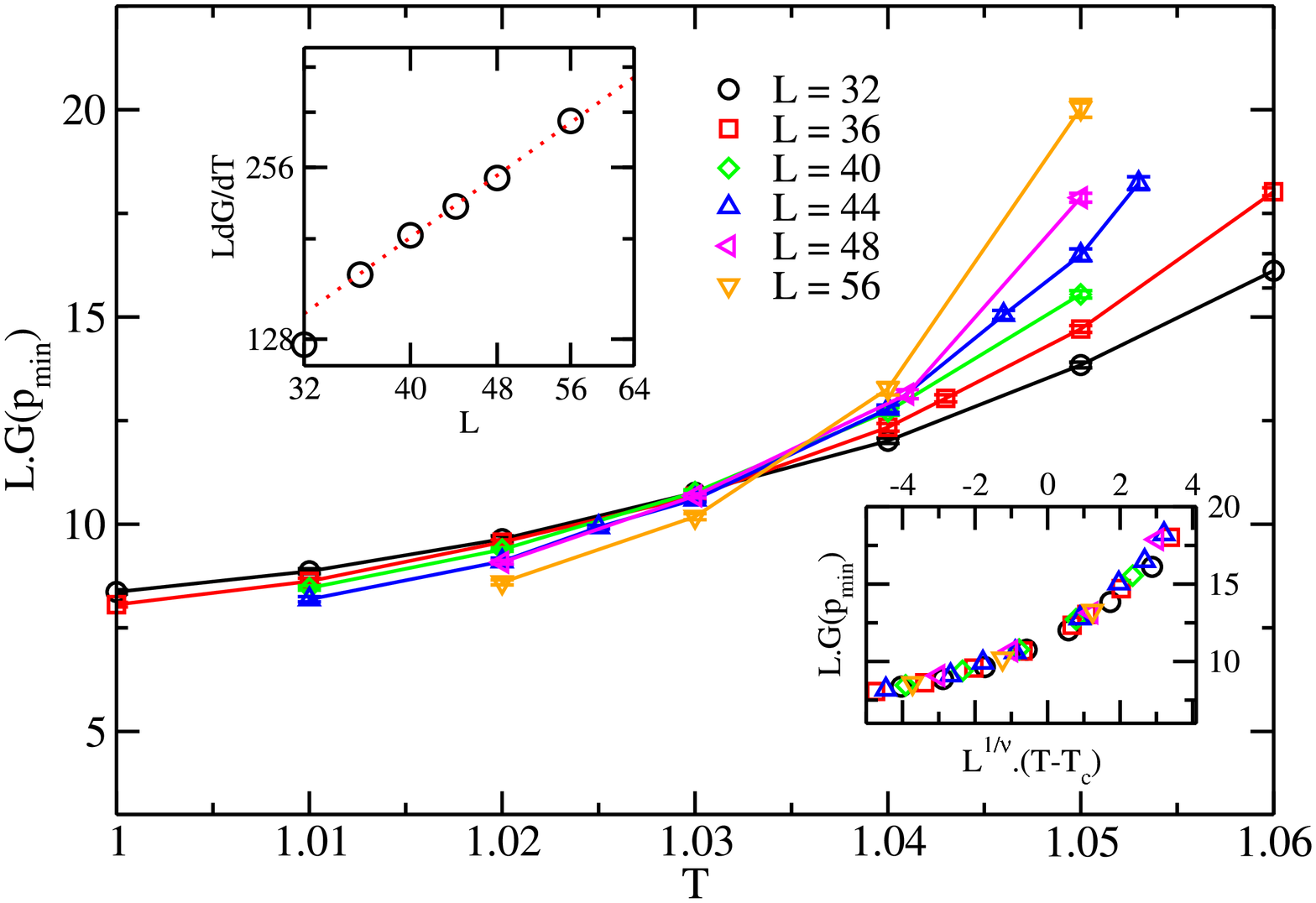}{\vspace*{-0.3cm}}
\caption{(color online) $L.G$ versus $T$ for different system sizes. Left
inset: Scaling of the derivative $LdG/dT$ versus system size $L$ (log-log
scale). Right inset: Data collapse of $L.G$.}
\label{fig:scaling}
\end{figure}

To characterize the high $T$ phase and detect its Coulomb nature, we study
the evolution of the transverse magnetic field correlator at small
momentum~\cite{Olsson, Kajantie}: 
$G(\textbf{p}_{\mathrm{min}}) = \frac{1}{L^3} \langle
\hat{B}_{i}(\textbf{p}_{\mathrm{min}})\hat{B}_{i}(-\textbf{p}_{\mathrm{min}})
\rangle$ with $\hat{B}_{i}(\textbf{p})$ the Fourier transform 
of $B_i(\textbf{r})$ and  $ \textbf{p}_{\mathrm{min}} = 2 \pi \hat{x}/L$. 
$G(\textbf{p}_{\mathrm{min}}) = 1$ if the gauge field presents dipolar
correlations and $G(\textbf{p}_{\mathrm{min}}) \to 0$ if the
field has short range correlations. In our case, we expect the field to 
acquire a mass in the low-T phase as the matter field condenses. The
evolution of $G(\textbf{p}_{\mathrm{min}})$ with $T$ is shown in
Fig.~\ref{fig:CV} b. At high $T$,  $G(\textbf{p}_{\mathrm{min}}) \to
1$ and the gauge field is gapless. As $T$ is lowered,
$G(\textbf{p}_{\mathrm{min}})$ decreases and the field becomes massive
due to the Higgs mechanism. Close to the critical point, we assume
the following scaling ansatz:  $L^{\eta_A}.G(\textbf{p}_{\mathrm{min}}) =
f(L^{1/\nu}(T-T_c))$, where gauge and scale invariance
impose $\eta_A = 1$ at the critical point~\cite{Tesanovic-Calan}. The
crossing of $L.G$ for different system sizes at $T_c$ has been observed
in the non-frustrated model, leading to an estimate $\nu =
0.67$~\cite{Olsson}. The frustrated model also presents a crossing point
(see Fig.~\ref{fig:scaling}) at $T_c^{G} = 1.035(5)$. The scaling versus
$L$ of the numerical derivative $LdG/dT$ leads
to an independent estimate $\nu=0.73(5)$ (see left inset of Fig.~\ref{fig:scaling}), agreeing with the
less precise values obtained from the Binder cumulant. The value of $\alpha$ obtained with hyperscaling  $\alpha = 2-
\nu d$ agrees with the convergence of $M_2$. The data collapse of $L.G$ obtained with this
value of $\nu$ is also of good quality (see right inset).

\begin{figure}
{\vspace*{-0.3cm}}\includegraphics[width=7cm]{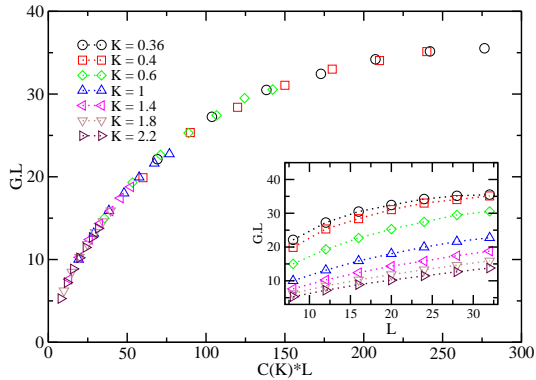}{\vspace*{-0.3cm}}
\caption{(color online) Data collapse of the Flowgram of the model. Inset: flows for different values of $\mathcal{K}$, note how the flows look more convergent for smaller stiffness.}
\label{fig:flowgram}
\end{figure}

So far, if the transition looks well continuous, we cannot totally exclude a very weak first-order process, with a very large but finite correlation length. 
In order to confirm this, we change the value of the stiffness $\mathcal{K}$ and implement 
the flowgram method~\cite{Kuklov,Kuklovnew}. This method relies on the demonstration that the large scale behavior for a large value of $\mathcal{K}$ is identical to that at a smaller 
value of $\mathcal{K}$ where the nature of the transition can be easily determined. 
For each $\mathcal{K}$, we follow Ref.~\onlinecite{Motrunichnew} and define the operational critical temperature 
	 $T_c(L)$ for a size $L$ as the temperature where the Binder ratio
is equal to $B_c = 1.41$~\cite{noteBC}. We then compute for each value of
$\mathcal{K}$ and $L$ the value of $L.G$ and recover the flows. If all the
curves within an interval $\mathcal{K} \in [\mathcal{K}_1,\mathcal{K}_2]$
can be  collapsed into a single master curve by rescaling the system size,
then it implies that the order of the transition remains the same within
the interval. A diverging flow for the collapse is a clear sign of a first
order transition. We present here the flowgram obtained for our model with
$\mathcal{K} \in [ 0.36,2.2 ]$ on Fig ~\ref{fig:flowgram}. We have
succeeded in performing a collapse of the curves by rescaling $L
\rightarrow C(\mathcal{K})L$ with $C(\mathcal{K}) = 2.0/\mathcal{K}
+0.4/\mathcal{K}^2$. The collapse is clearly \textit{converging}. This
suggests that scale invariance is reached and goes in favor of a continuous
transition in the interval considered, confirming the previous analysis at $\mathcal{K}=1$.

To summarize our results, we have shown how, via a duality transformation,
the interacting dimer model on the cubic lattice can be understood in terms
of a complex matter field coupled to a gauge field with geometric
frustration. As for the microscopic dimer model~\cite{Alet}, we found a
direct continuous transition  (by the Higgs mechanism) between a dipolar
Coulomb phase at low coupling and a columnar phase which breaks lattice symmetries at high coupling. However, critical exponents are rather different from the ones obtained in the
original dimer model (where $\alpha \simeq 0.5$ and $\nu \simeq 0.5$). In
particular, the second moment of the action clearly shows no diverging peak
in our case. The anomalous dimension $\eta_A$ is in agreement with the scale dimension of the stiffness
in Ref.~\onlinecite{Alet} but a different value of $\nu$ is obtained. 
The exponents that we find ($\nu \simeq 0.73$ and $\alpha < 0$) seem
to be slightly different from those of the 3d XY universality class (and
consequently of the non-frustrated version
of Eq.~\ref{lattice}), although we cannot exclude it within our numerical
accuracy. In both cases, we are left with an interesting open problem as
there is no field theoretical arguments allowing to say that we should end up in the
XY universality class. In view of our
original motivation, an intriguing possibility remains that the microscopic dimer model is
{\it directly} self-tuned (for unknown reasons) to a tricritical
point.  The flowgram analysis displays no sign of tricriticality however in the range of stiffness considered.  Since values of $\mathcal{K} < 0.36$ cannot be reached as the critical temperature is too low for the Metropolis algorithm to be efficient, one alternative would be to directly perturb the original
microscopic dimer model to check this scenario. 

We thank F. Delduc, I. Herbut, A. Honecker, G. Misguich, V. Pasquier and A. Vishwanath
for enlightening discussions, and GENCI for allocation of CPU time. Simulations used the ALPS libraries~\cite{ALPS}.


\begin{thebibliography}{99}
\bibitem{noteBC} We checked that the results hold if $B_c$ is slightly varied.
\end{thebibliography}
\end{document}